%% file: paper.tex
\documentclass[12pt]{article}
\usepackage{a4wide}
\usepackage{epsfig}

\newlength{\figwidth}
\usepackage{mcite}
\usepackage{rotating}

\newcommand{\new}{\newcommand}

\def\sinc{\ensuremath{\sigma_{\mathrm{inc}}}}
\def\sacc{\ensuremath{\sigma_{\mathrm{acc}}}}
\def\H{\ensuremath{\mathrm{H}\;}}

\def\gg{\ensuremath{\gamma\gamma}}
\def\ggpX{\ensuremath{\gamma\gamma+X}}
\def\pp{\ensuremath{\mathrm{pp}\;}}
\def\ra{\ensuremath{\rightarrow}}
\def\mh{\ensuremath{m_\mathrm{H}}}

\def\gev{\ensuremath{\;\mathrm{GeV}}}
\def\gevoc{\ensuremath{\;\mathrm{GeV}/c}}
\def\gevc2{\ensuremath{\;\mathrm{GeV}/c^2}}
\def\frag{\ensuremath{\mu_{\mathrm{F}}\;}}
\def\renorm{\ensuremath{\mu_{\mathrm{R}}\;}}
\def\ptm{\ensuremath{p_{\mathrm{T}}^{\mathrm{m}}}}
\def\ys{\ensuremath{Y^*}}
\def\pt{\ensuremath{p_{\mathrm{T}}}}

\new{\VEGAS}        {\mbox{\small\textsc{VEGAS}}}
\new{\MCaNLO}        {\mbox{\small\textsc{MC@NLO}}}
\new{\FEHiP}            {\mbox{\small\textsc{FEHIP}}}
\new{\DiPHOX}         {\mbox{\small\textsc{DIPHOX}}}
\new{\MRST}        {\mbox{\small\textsc{MRST01}}}
\new{\CTEQ}        {\mbox{\small\textsc{CTEQ6M}}}
\new{\CERN}        {\mbox{\small\textsc{CERN}}}
\new{\LHC}        {\mbox{\small\textsc{LHC}}}
\new{\ATLAS}        {\mbox{\small\textsc{ATLAS}}}
\new{\CMS}        {\mbox{\small\textsc{CMS}}}
\new{\HERWIG}        {\mbox{\small\textsc{HERWIG}}}
\new{\HDECAY}        {\mbox{\small\textsc{HDECAY}}}
\new{\HIGLU}        {\mbox{\small\textsc{HIGLU}}}

\begin{document}


\begin{flushright}
        ETHZ-IPP-PR-2005-02\\
        \today
\end{flushright}

\vspace{1cm}
\begin{center}

  {\Huge{Study of perturbative QCD predictions at\\[-0.1cm] 
  next-to-leading order and beyond\\[0.3cm]
  for \pp \ra \H \ra \ggpX}}

\vspace{1cm}

  {\large Fabian St\"ockli, Andr\'e G. Holzner, G\"unther Dissertori}\\[0.3cm]
          {\small Institute for Particle Physics,
                        ETH Zurich, Switzerland}

\vspace{1cm}

  \begin{abstract}
    \input{abstract.tex}
  \end{abstract} 

\end{center}

\newpage

\section{Introduction}
\input{introduction.tex}


\section{Computer programs and parameter settings}\label{sec:programs}
\input{programs.tex}
\section{Event selection}\label{sec:selection}
\input{selection.tex}


\section{Inclusive and accepted cross sections}\label{sec:incxsect}
\input{inc_cross.tex}

\section{Differential cross sections}\label{sec:diffxsect}
\input{diff_cross.tex}

\section{Comparison to the irreducible background}\label{sec:background}
\input{background.tex}

\section{Summary}\label{sec:summary}
\input{summary.tex}

\section*{Acknowledgements}

We would like to warmly thank K.\ Melnikov, F.\ Petriello and in particular
C.\ Anastasiou for the many and very
interesting discussions, for the help with the \FEHiP\ code and the comments
on the manuscript. Furthermore we are grateful to S.\ Frixione and
B.\ Webber for their feedback on our questions related to
\MCaNLO, as well as for the useful comments on the
manuscript. We thank M.\ Spira for his advice on theoretical issues.

\bibliographystyle{utphysown}
\bibliography{paper}

\end{document}

%% file: abstract.tex
We study predictions from perturbative Quantum Chromodynamics (QCD) for the process \pp \ra \H \ra \ggpX. 
In particular, we compare fully differential calcu\-lations at next-to-leading  (NLO) and 
next-to-next-to-leading order (NNLO) in the strong coupling constant to the results obtained with the \MCaNLO\ Monte Carlo (MC) generator, which combines 
 QCD matrix elements at NLO with a parton shower algorithm.  Estimates for the systematic uncertainties in the various predictions 
due to the choice of the
renormalization scale and the parton distribution functions are given for the inclusive and accepted cross sections and for the corresponding 
acceptance corrections,
obtained after applying standard selection and acceptance cuts.
Furthermore, we compare the distributions for the Higgs signal to those  for the irreducible two-photon background, obtained with a NLO MC simulation.


%% file: introduction.tex
One of the main goals of the future Large Hadron Collider (\LHC) at
\CERN\ is the search for the Higgs boson. 
From theoretical considerations \cite{Hambye:1997ax} it is concluded that the 
Higgs mass should be below about one TeV/$c^2$. 
In addition, precision measurements
of electro-weak observables \cite{Altarelli:2004fq} indicate that the low-mass  
region (approximately below 200\gev/$c^2$) is preferred.
If the Higgs boson mass is smaller than twice the W mass,
its decay mode into a pair of photons will be among the most
important and distinct channels for the Higgs search. Since this decay
can only proceed via loop corrections, the corresponding branching ratio
is less than one percent. However, it provides the
experimentally cleanest signature~\cite{unknown:1999fr,:1997kj,Abdullin:2005yn}.

Obviously, one important aspect of the preparations for the upcoming \LHC\ Higgs searches
is to study the predictions for the  distributions of
kinematic variables for the signal, $\pp \ra \H \ra \ggpX$, and the background processes (eg.\
$\pp \ra  \ggpX$), obtained from  perturbative Quantum Chromodynamics
(QCD). 
First, 
it is known that the inclusive Higgs cross section  gets large perturbative corrections at
next-to-leading order (NLO) \cite{Dawson:1990zj,Djouadi:1991tk,Graudenz:1992pv,Spira:1995rr}
and next-to-next-to-leading order (NNLO)
\cite{Catani:2001ic,Harlander:2001is,Harlander:2002wh,Anastasiou:2002yz,Ravindran:2003um}, 
so it is likely that the dynamics of the process
is different at various orders in perturbation theory. 
That is, the QCD corrections to the
production process change the kinematics of the Higgs boson and lead to modifications
in the kinematics of the final-state photons. It is worth noting here that
very recently the leading threshold-enhanced third-order
 ($\mathrm{N}^3$LO) corrections to the total Higgs cross sections have
 been derived \cite{Moch:2005ky}, showing that the perturbative expansion stabilizes.

Second, the detailed knowledge of differential cross sections will help to optimize the
search strategy and thus improve the signal over background ratio. In the past,
such studies have been performed using leading order (LO) Monte Carlo
generators, which usually include a parton shower and a hadronization model and 
are finally combined with  a full detector simulation. Higher order
QCD predictions in general were only available for the inclusive cross sections or for
distributions of some kinematic variable of the Higgs boson, for example its transverse
momentum, \pt~\cite{deFlorian:1999zd,Ravindran:2002dc}.  Therefore, the
$K$-factors applied to the LO predictions to correct for the missing higher order terms
would not take account of the analysis selection and acceptance cuts, which might reduce the
 allowed phase space in a non-trivial way. An attempt to improve on this consists in 
 a re-weighting procedure where the LO plus parton shower results are re-scaled in
 order to reproduce the higher order prediction for a particular
 kinematic variable, such as the Higgs \pt. Taking account of these weights
and of the efficiencies obtained from the MC simulation, an effective $K$-factor can be
computed which incorporates the selection and acceptance factors \cite{Davatz:2004zg}.

Recently, new calculations for the  process \pp \ra \H \ra \ggpX\ beyond  
NLO  have become available, which
now allow for an analysis of acceptance corrections and the related systematic
uncertainties at  higher order in perturbative QCD. In particular,  the fully 
differential cross section at NNLO has been
computed for the first time \cite{Anastasiou:2004xq,Anastasiou:2005qj}, giving
the exact expressions (up to this order of perturbative QCD)
for Higgs production and its two-photon decay with up to
two additional massless partons (quark or gluons). Therefore this new semi-analytical calculation allows
to apply realistic selection and isolation cuts to the Higgs decay products and the additional
partons in the final state, as well as to study the distribution of relevant kinematic variables,
such as the rapidity and \pt\ of the Higgs and the photons, at NNLO accuracy.  

An alternative approach consists in the combination  
of QCD matrix elements at NLO and a parton shower algorithm, therefore
taking into account also large logarithms due to soft and collinear radiation at all orders in the
strong coupling constant. This is implemented in the Monte
Carlo (MC) generator \MCaNLO\ \cite{Frixione:2002ik,Frixione:2003ei}, combined with the
\HERWIG\ MC program \cite{Corcella:2000bw,Corcella:2002jc}, which in addition handles the hadronization process and 
particle decays. Several aspects render such an approach interesting for phenomenological 
studies and experimental analyses. In contrast to a pure parton level calculation, such a MC 
generator allows to investigate distributions at the particle (hadrons and leptons) level, which is
closer to the actually measured final state than partons.  For example, the detailed internal structure
of jets is modeled rather well,  studies of lepton isolation from jets are expected to be more
realistic and it is possible to simulate also the effects of the underlying event and multiple
interactions. Finally, it is rather straightforward to feed the output of this generator into
a full detector simulation, which is necessary for the experimenter in order to develop  optimised
data analyses which rely on a good approximation of the underlying QCD dynamics. 
There is already a considerable list of processes available in \MCaNLO, among
these \pp \ra \H \ra \ggpX, and further additions can be expected in the future.

This paper provides a first detailed comparison of these newly available predictions. 
Taking the various calculations, 
we apply standard  cuts to relevant kinematic observables in order to reproduce in a realistic manner the basic
aspects of the Higgs search  in the two-photon channel at \LHC. As a result, we are
able to compare the predictions for the inclusive and accepted cross
sections, as well as for the corresponding acceptance
corrections. These comparisons are done as a function of the Higgs
mass and the 
renormalization/factorization scales. In addition, we study the dependence of the \MCaNLO\ predictions on the
choice of the parton distribution functions (pdf). Regarding the latter, it can be interesting to compare
the predictions when combining \MCaNLO\ with pdfs extracted at NLO as well as at NNLO accuracy.
Although via the combination of NLO matrix elements with the initial state parton shower effectively 
also terms beyond NLO are generated for the perturbative part of the scattering process,
the use of NNLO pdfs in \MCaNLO\ should be regarded as equivalent to that of NLO pdfs,
since, formally, the accuracy of the results is NLO in both cases.

Interesting kinematic observables of the two-photon
final state, as proposed in \cite{Bern:2002jx} and \cite{Anastasiou:2005qj}, are computed for the signal 
as well as  for the irreducible two-photon background, \pp \ra \ggpX. 
The latter is known at NLO accuracy \cite{Bern:2002jx, Binoth:1999qq}. 
It is shown that variables such as the average photon transverse momentum or the rapidity difference
of the two photons could contribute to the discrimination of the signal from the background.

It is worth noting that the entire study is done at the phenomenological level, i.e.\ no
attempt is made here to simulate any detector effects beyond the basic selection criteria, such as 
acceptance cuts in photon rapidity and transverse momentum, as well as isolation requirements. 
We thus emphasize that this paper does not provide a re-evaluation of the Higgs
discovery potential at \LHC,  but rather intends to show the level of theoretical 
accuracy, which is achievable with the tools that have become available recently, as well as to
indicate where more detailed studies, including full detector simulation and the most up-to-date
perturbative QCD predictions, might  be worthwhile pursuing.  


The paper is organized as follows: 
In Section~\ref{sec:programs} we list the details on the computer programs and the
relevant parameter settings which were 
used to perform this study.
In Section~\ref{sec:selection} the basic event selection cuts are described.
The resulting cross sections before and after this selection are discussed 
 in Section~\ref{sec:incxsect} and the distributions of some kinematic
variables are presented in Section~\ref{sec:diffxsect}. The comparison
to the main background is given in Section~\ref{sec:background}.
Finally, Section~\ref{sec:summary} contains a summary of the results. 


%% file: programs.tex
In order to carry   out this  study  we  have used  several  different
software packages, which implement the  various signal and  background
predictions up to a  certain  level of approximation in   perturbation
theory. The employed programs are~:

\begin{itemize}
\item \FEHiP\ \cite{Anastasiou:2005qj} (Version 1.0)\\
  This code is used to calculate the fully differential cross section
  for Higgs boson production through gluon  fusion at NLO and NNLO
  QCD, in the  narrow-width approximation.   The gluon fusion  channel
  via a  top quark loop  contributes a fraction of  about  80\% to the
  total  Higgs production   cross section  in  the relevant   low-mass
  region.  The  calculation has  been carried  out in  the limit of an
  infinitely heavy  top quark,  in which  the Higgs boson  coupling to
  gluons is point-like. However,  the results are re-scaled  using the
  ratio of  the LO cross section predictions,  obtained with the exact
  top mass dependence and in the heavy top mass limit. As discussed in
  \cite{Anastasiou:2005qj}, this gives an excellent approximation.

For  the consistent convolution of the  partonic cross sections at NLO
(NNLO) with  parton   distribution functions,  the \MRST\  NLO  (NNLO)
\cite{Martin:2002dr} pdf sets  are used. The so-called \textsc{mode 1}
is  employed as the default  mode  for pdf  evolution.  For the moment
being,  no other pdf set is  available with this   program and we have
made no attempt to include further sets.

The  code returns the  four-momenta of the  Higgs boson and the final state
particles,  namely two  photons   and up  to  two additional  massless
partons, which can be used to evaluate an  analysis function for every
event.  This  analysis function is  then  convoluted  with the squared
matrix elements  of the corresponding  final state in order to finally
obtain a cross section.  Many of  the numerical results, in particular
at NNLO, were obtained from
\cite{Babis-Kyrill-Frank-private-communication,Anastasiou:2005qj}. 
Concerning statistical uncertainties, the NNLO  results for the  cross
sections (acceptance corrections) are accurate to 1\% (2\%), while the
NLO numbers have been computed up to an accuracy of 0.1\%. 

In order to speed up the computation, for the Monte Carlo integration 
step a parallel \VEGAS\ algorithm has been used \cite{kreckel-1997-106,kreckel-1998-127}. 
The available C++ code has been slightly modified in order to meet the specific needs of \FEHiP.
The code has been parallelized using the Standard Message Passing Interface MPI \cite{mpi}. 

\item \MCaNLO\ \cite{Frixione:2002ik,Frixione:2003ei} (Version 2.31)\\
  As described in the   introduction, this MC generator   consistently
  incorporates NLO QCD matrix elements into a parton shower framework.
  In the purely inclusive case the code returns the NLO cross section, 
  whereas the effects of the parton shower
  become only apparent when cuts are applied to the final state.
  In order to eliminate a problem with this version of the code
  when using renormalization and factorization scales that differ from 
  the default choice ($\mu=\mh$), a correction has been applied \cite{Frixione:private-communication}.
  
  \MCaNLO\ is  used    together  with    the   \HERWIG\  event     generator
  \cite{Corcella:2000bw,Corcella:2002jc}, which handles the simulation
  of the parton shower, the hadronization   process   and possible
  particle decays. The exact top  mass dependence is retained at  the
  Born  level.  We  have generated  events  using three different  pdf
  sets. In  addition to the two \MRST\  sets mentioned above, we have also
  used the \CTEQ\ \cite{Pumplin:2002vw}  set, which has been extracted
  at NLO accuracy.

\item \DiPHOX\ \cite{Binoth:1999qq} (Version 1.2)\\
  The background process \pp \ra \ggpX\ is computed with this program,
  which provides a NLO  QCD prediction at  parton level only. 
  The (unphysical) phase-space slicing parameter $p_{\mathrm{T}_\mathrm{m}}$~\cite{Binoth:1999qq}
was set to $0.05 \gevoc$.
  An additional contribution to this process, which is not included in \DiPHOX, 
  namely the two-loop correction to the gluon fusion subprocess,
  has been calculated in
  \cite{Bern:2002jx}.  There it  is  shown that this contribution   is
  modest, around 10\% or smaller for the mass range of interest in the
  low-mass Higgs search.

\item \HDECAY\ \cite{Djouadi:1997yw} (Version 3.101)\\
  This is used to calculate the branching ratio  of the Higgs decay to
  two photons, resulting in BR(\H \ra \gg) = 0.22 \%.

\item \HIGLU\ \cite{Spira:1995mt} (Version 2.102)\\
 The program \HIGLU\ allows to calculate the inclusive cross section
  for Higgs boson production through gluon  fusion at NLO, with the exact 
  top mass  dependence at both LO and NLO.

\end{itemize}

Throughout this analysis a top mass of $m_\mathrm{top} = 175 \gev/c^2$
has been  assumed.  Whenever we refer  to `the scale' $\mu$,  it means
$\mu = \mu_{\mathrm{R}}=\mu_{\mathrm{F}}$, i.e.\ the factorization and
renormalization scales are set to the same value.


%% file: selection.tex
In this study we attempt to simulate in an appropriate manner
the planned Higgs searches at \LHC\ in the two-photon channel,
using only phenomenological, i.e.\ parton and/or particle level, predictions without any 
detailed detector simulation. This is achieved by
taking into account the geometrical limitations of a typical \LHC\ detector and
the foreseen trigger requirements. Furthermore, in order to  
reduce the large irreducible background the following cuts are applied to the
final state photons~: 

\begin{itemize}
\item The absolute value of the pseudorapidity
  $|\eta_{\gamma_{1,2}}|$ of each of the two photons must be smaller
  than 2.5. This corresponds to the coverage in polar angle of a
  detector such as \ATLAS~\cite{unknown:1992ee} or \CMS~\cite{DellaNegra:1992hp}. 

\item One of the two photons must have a transverse energy above 40\gev,
  while the other photon is required to have a transverse energy above 25\gev.

\item In order to exclude events with considerable hadronic activity around the final state photons, as for example can be expected in background events with quark-photon fragmentation, the photons must be isolated, i.e.\ the additional energy in a cone
  $\Delta R = \sqrt{\Delta \eta^2 + \Delta \varphi^2} < 0.4$ around 
  each photon must not exceed 15 GeV. 

\end{itemize}

This selection is basically the same as the `standard cone criterion'
used in~\cite{Bern:2002jx}. For reasons given in that reference
(e.g.\ calorimeter granularity and finite photon shower size), we
did not investigate the smooth cone isolation criterion \cite{Frixione:1998jh}.
In the following, we will refer to these selection cuts as the `standard cuts'
and call the accepted photon with the highest (second highest)
transverse energy photon~1 (photon~2).


%% file: inc_cross.tex
In the following section we study the predictions at various orders
in perturbative QCD for  the inclusive \pp\ra\H\ra\ggpX \
cross section \sinc, as well as for the accepted cross section \sacc\ found after the standard cuts 
on the final state photons have been applied. The comparison is made for several
different choices of parameters, such as the Higgs mass, the scale or the pdf set.  
For the moment we simply assume that the branching ratio of the Higgs decay into two 
photons is one. 

In a first step, we fix the Higgs mass to $\mh=120\gevc2$, the
factorization scale \frag{} and renormalization scale \renorm{}
are set to \mh/2 and the pdfs are taken from the \MRST\ set, using the NLO (NNLO) pdfs in combination
with \MCaNLO\ and \FEHiP\ at NLO (NNLO). As discussed in detail in Ref.\ \cite{Anastasiou:2005qj}, 
the choice $\frag{}=\renorm{}=\mh/2$ is motivated by the observation of an improved convergence
of the perturbation series and the very good agreement with the 
threshold-resummed results
for the Higgs hadroproduction cross section \cite{Catani:2003zt}. This is further
confirmed by the recent calculation of the threshold-enhanced third-order
corrections \cite{Moch:2005ky}.

The results can be found in  Table~\ref{tab:higgs-efficiencies-cuts}. Whereas with \MCaNLO\
we give the acceptances for each single cut as well as for groups of cuts, in the case of 
\FEHiP\ we had to restrict ourselves to the acceptance after all cuts, since this NNLO code is
very cpu-intensive.
Regarding the inclusive cross section, \sinc, we observe that the \MCaNLO\ prediction falls short of the 
NLO prediction of \FEHiP\ by about 3.3\%. By construction \MCaNLO\ should
return the exact NLO cross section if no cuts are applied. Thus the
observed difference has to be explained by the different treatment of the
top mass dependence. \FEHiP\ uses the infinite top mass limit in all orders of the calculation and then re-scales the result using the ratio of the
exact and approximated ($m_\mathrm{top}\rightarrow\infty$) LO cross sections, while \MCaNLO\ keeps the exact top mass dependence at LO, but uses the $m_\mathrm{top}\rightarrow\infty$ approximation at NLO. For comparison, the
result obtained with \HIGLU\ which retains the exact top mass dependence
at LO and NLO is $\sigma_{\mathrm{inc}} = 44.39 \pm 0.01$ pb. Thus 
\FEHiP\ returns a 0.9\% higher and \MCaNLO\ a 2.5\% 
lower NLO cross section with respect to \HIGLU.

The differences of the NLO results with respect to the NNLO result amount to 13\% (\MCaNLO) and 
9\% (\FEHiP, NLO). After the
standard selection, these differences are reduced to about 10\% and 6\%, respectively. 
The acceptance corrections for $\mh=120\gevc2$, defined as the ratio between the accepted and the inclusive
cross sections, $\sacc/\sinc$, agree to within 2\% (absolute).

 \begin{sidewaystable}[htbp]
 \centering
  \begin{tabular}{|l||c|c|c||c||c|} \hline\hline
    Generator                           & \multicolumn{3}{c||}{\MCaNLO}             & \FEHiP\ NLO     & \FEHiP\ NNLO     \\ \hline\hline  
    PDF set                             & \MRST\ NLO    & \MRST\ NNLO    & \CTEQ\     & \MRST\ NLO      & \MRST\ NNLO    \\ \hline  
    $\sigma_{\mathrm{inc}}$ [pb]        & 43.30         & 37.67          & 44.32      &  44.77          &   48.94        \\ \hline   
    \multicolumn{6}{c}{}                                                                                                 \\ \hline
    \multicolumn{6}{|c|}{Acceptances ($\sigma_{\mathrm{acc}}/\sigma_{\mathrm{inc}}$) of single groups of cuts}           \\ \hline
    $p_{\mathrm{T}}$-cuts               & 80.2 \%       & 80.3 \%        & 80.1 \%    &                 &                \\ 
    $\eta$-cuts                         & 83.2 \%       & 83.0 \%        & 83.8 \%    &                 &                \\ 
    isolation                           & 99.7 \%       & 99.7 \%        & 99.8 \%    &                 &                \\ \hline

    \multicolumn{6}{c}{}                                                                                                 \\ \hline
    \multicolumn{6}{|c|}{Acceptances ($\sigma_{\mathrm{acc}}/\sigma_{\mathrm{inc}}$) of two groups of cuts}              \\ \hline
    $p_{\mathrm{T}}$- and $\eta$-cuts   & 63.6 \%       & 63.2 \%        & 64.3 \%    &                 &                \\ 
    $p_{\mathrm{T}}$-cuts and isolation & 79.6 \%       & 79.7 \%        & 79.6 \%    &                 &                \\ 
    $\eta$-cuts and  isolation          & 81.5 \%       & 81.1 \%        & 82.1 \%    &                 &                \\ \hline 
    \multicolumn{6}{c}{}                                                                                                 \\ \hline
    \multicolumn{6}{|c|}{All three groups of cuts}                                                                       \\ \hline
    Acceptance                          & 63.0 \%       & 62.6 \%        & 63.7 \%    & 63.4 \%         & 61.4 \%        \\ \hline
    $\sigma_{\mathrm{acc}}$ [pb]        & 27.30         & 23.60          & 28.25      & 28.35           & 30.07          \\ \hline    
  \end{tabular}
  \caption{Cross sections and acceptance corrections for a Higgs mass of $\mh=120 \gevc2$, 
   obtained with \MCaNLO\ and \FEHiP\ (NLO and NNLO)
    for different cuts and pdf sets. The renormalization and factorization scale are
     set to $\mu=\mh/2$. The branching ratio \H\ra\gg\ is assumed to be one.   
     The statistical uncertainties are 
    at the one (two) per-cent level for the NNLO cross sections (acceptance)
    and negligible for the other cases.
    \label{tab:higgs-efficiencies-cuts}}
\end{sidewaystable}

It turns out that the NNLO calculation
predicts a slightly stronger reduction in the accepted cross section than the other two
approaches. With the \MCaNLO\ generator we have also studied the effect of the individual
cuts. We find that the reduction in cross section is mainly due to the angular and momentum cuts,
whereas the isolation requirement has only a very minor impact.


\subsection{Choice of pdfs in \MCaNLO}
 \label{pdf-inc}

As a first variation with respect to the parameter choices given above, we investigate the 
dependence on the pdfs used together with \MCaNLO. 
Since \MCaNLO\ includes the leading effects of soft and collinear radiation
to all orders, thus beyond NLO, it is interesting to  combine it with the NNLO pdf set of the
\MRST\ fits, which has been extracted using a combination of NLO and NNLO cross sections.  
Finally, we also take \CTEQ\ (NLO) as input, 
in order to study an alternative set of  pdfs, obtained under different assumptions, from a somewhat
different data set and thus with different theoretical and experimental
systematic uncertainties.
 
The results are listed in Table~\ref{tab:higgs-efficiencies-cuts}. Compared to \MRST\ NLO,
the results  found with the \MRST\ NNLO set, for both \sinc\ and \sacc, differ quite considerably, 
up to 16\%, whereas the cross sections obtained with the NLO sets from the \MRST\ and \CTEQ\ fits agree
to better than 4\%. The considerably lower cross sections obtained with the NNLO set are interpreted
as arising from the smaller gluon density and strong coupling used in this set. It is worth noting that
in the NNLO calculation, as implemented in \FEHiP, the NNLO corrections in  the matrix elements,
which also include additional scale dependent terms,  counter the effect of the NNLO evolution of the pdfs.

Interestingly, when looking at the acceptance, we find agreement for all 
three choices within 1.1\%, absolute; when also considering the 
\FEHiP\ results the span between the largest and smallest acceptance 
increases to 2.3\%.  Half of this span could be defined as  a 1.8\%  `uncertainty' relative to
the average of the largest and smallest values. Calculating 
the same relative uncertainty for the accepted cross section, we find  4.7\%. Thus 
the acceptance correction is less sensitive with respect to 
the choice of pdfs and the approximations made by the perturbative calculation
 than the absolute cross section.


\subsection{Variation of the Higgs mass}

In Table~\ref{tab:higgs_mass_cuts} the results are given for the inclusive and the
accepted cross sections for different Higgs masses, obtained with
\MCaNLO\ and \FEHiP\ (NNLO). The renormalization and
factorization scales are chosen to be $\mu=\mh/2$. 

\begin{sidewaystable}[htbp]
  \centering
\resizebox{!}{4cm}{
  \begin{tabular}{|l||c|c||c|c||c|c||c|c||c|c|} \hline
    Generator                           & \MCaNLO & NNLO    & \MCaNLO & NNLO    & \MCaNLO & NNLO    & \MCaNLO & NNLO    & \MCaNLO & NNLO       \\ \hline
    $\mh$ [GeV]& \multicolumn{2}{|c||}{$110$} &\multicolumn{2}{|c||}{$120$}   &\multicolumn{2}{|c||}{$130$}  &\multicolumn{2}{|c||}{$140$}  & \multicolumn{2}{|c|}{$150$}  \\ \hline
    \sinc [pb]                          & 50.65   & 56.82   & 43.30   & 48.94   & 37.45   & 42.54   & 32.66   & 37.37   & 28.70   & 33.10      \\ \hline
    \multicolumn{11}{c}{}                                                                                                                      \\ \hline
    \multicolumn{11}{|c|}{Acceptances (\sacc/\sinc) of single groups of cuts}                                                                  \\ \hline
    $p_{\mathrm{T}}$-cuts               & 75.9 \% &         & 80.2 \% &         & 83.3 \% &         & 85.9 \% &         & 87.8 \% &            \\ 
    $\eta$-cuts                         & 82.4 \% &         & 83.2 \% &         & 84.0 \% &         & 84.8 \% &         & 85.3 \% &            \\ 
    isolation                           & 99.7 \% &         & 99.7 \% &         & 99.7 \% &         & 99.7 \% &         & 99.7 \% &            \\ \hline
    \multicolumn{11}{c}{}                                                                                                                      \\ \hline
    \multicolumn{11}{|c|}{Acceptances (\sacc/\sinc) of two groups of cuts}                                                                     \\ \hline
    $p_{\mathrm{T}}$- and $\eta$-cuts   & 59.9 \% &         & 63.6 \% &         & 66.3 \% &         & 68.6 \% &         & 70.5 \% &            \\
    $p_{\mathrm{T}}$-cuts and isolation & 75.4 \% &         & 79.6 \% &         & 82.7 \% &         & 85.2 \% &         & 87.1 \% &            \\
    $\eta$-cuts and  isolation          & 80.7 \% &         & 81.5 \% &         & 82.1 \% &         & 82.8 \% &         & 83.3 \% &            \\ \hline
    \multicolumn{11}{c}{}                                                                                                                      \\ \hline
    \multicolumn{11}{|c|}{All three groups of cuts}                                                                                            \\ \hline
    Acceptance                          & 59.5 \% & 57.3 \% & 63.0 \% & 61.4 \% & 65.7 \% & 63.7 \% & 67.9 \% & 64.7 \% & 69.7 \% & 67.4 \%    \\ \hline
    $\sigma_{\mathrm{acc}}$ [pb]        & 30.11   & 32.56   & 27.30   & 30.07   & 24.60   & 27.08   & 22.19   & 24.17   & 20.01   & 22.30      \\ \hline
  \end{tabular}
}
\caption{Acceptance corrections for the different cuts and the different Higgs masses, as obtained with \MCaNLO\ and 
             \FEHiP\ (NNLO).  The used pdfs are \MRST\ NLO for \MCaNLO\ and \MRST\ NNLO for \FEHiP. The renormalization and factorization scale are set to $\mu=\mh/2$. 
  The statistical uncertainties are 
  at the one (two) per-cent level for the NNLO cross sections (acceptance)
  and negligible for the other cases.}
\label{tab:higgs_mass_cuts}
\end{sidewaystable}

In the upper part of the table the inclusive cross sections \sinc\ for \MCaNLO\ (\MRST\ NLO)
and the NNLO results obtained with \FEHiP\ (\MRST\ NNLO) are compared.
It can be observed that the inclusive cross sections obtained with 
\MCaNLO\ are  $16\%$ smaller than the NNLO results for the full Higgs mass range
investigated here. In the second part of Table~\ref{tab:higgs_mass_cuts} the 
\MCaNLO\ predictions for the acceptance corrections, after the various cuts, are given.
As stated above, the isolation cut has the smallest effect on the
acceptance corrections. The increased acceptance of the cuts on $p_{\mathrm{T}}$ and
$\eta$ when increasing the Higgs mass arises from the higher 
transverse momenta expected in this region. Therefore also the total acceptance
increases with larger \mh. The results after cuts, obtained with \MCaNLO\ and \FEHiP,
agree to within 11\%. This reduced difference compared to the inclusive case
 is due to the slightly larger acceptance correction found with \MCaNLO.
Remarkably, the acceptance corrections 
for \MCaNLO\ and \FEHiP\ (NNLO) agree within 3.2\% (absolute) over the whole investigated mass range.
 
\begin{figure}[htb]
  \centering
  \includegraphics[scale=0.5]{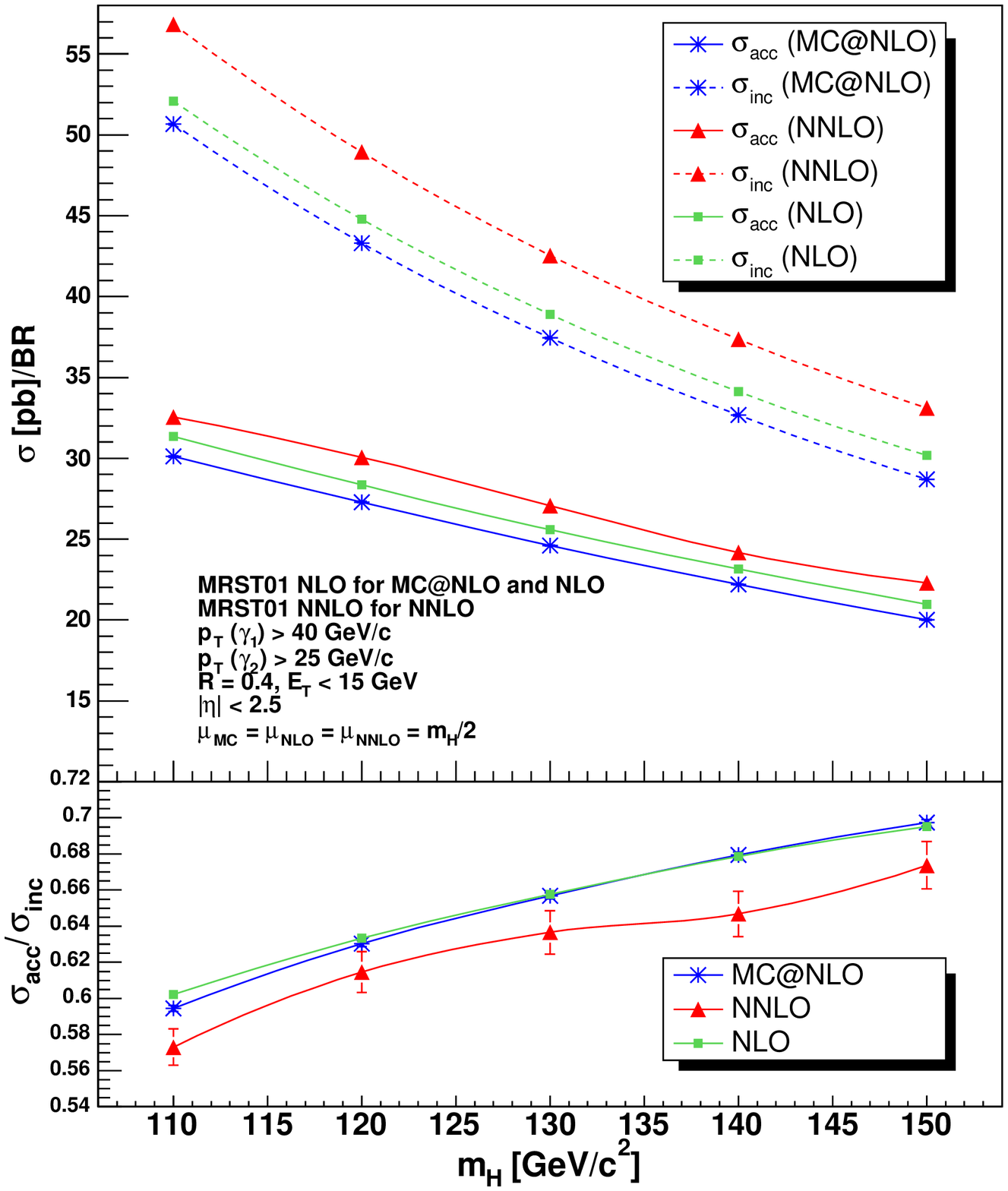}
  \caption{Inclusive and accepted cross sections as a function of the
    Higgs mass, \mh, obtained with \MCaNLO\ and \FEHiP\ (at NLO and NNLO).
    \label{fig:acc_cross_mh}
  }
\end{figure}

The results are shown in Fig.~\ref{fig:acc_cross_mh}, where the inclusive
and accepted cross section, as well as the acceptance correction, are plotted as a function of \mh.
It is worth noting that the statistical uncertainty of the NNLO results for the acceptance correction
is at the 2\% level.

\subsection{Variation of the renormalization and factorization scale $\mu$}

\begin{figure}[htb]
  \centering
  \includegraphics[scale=0.5]{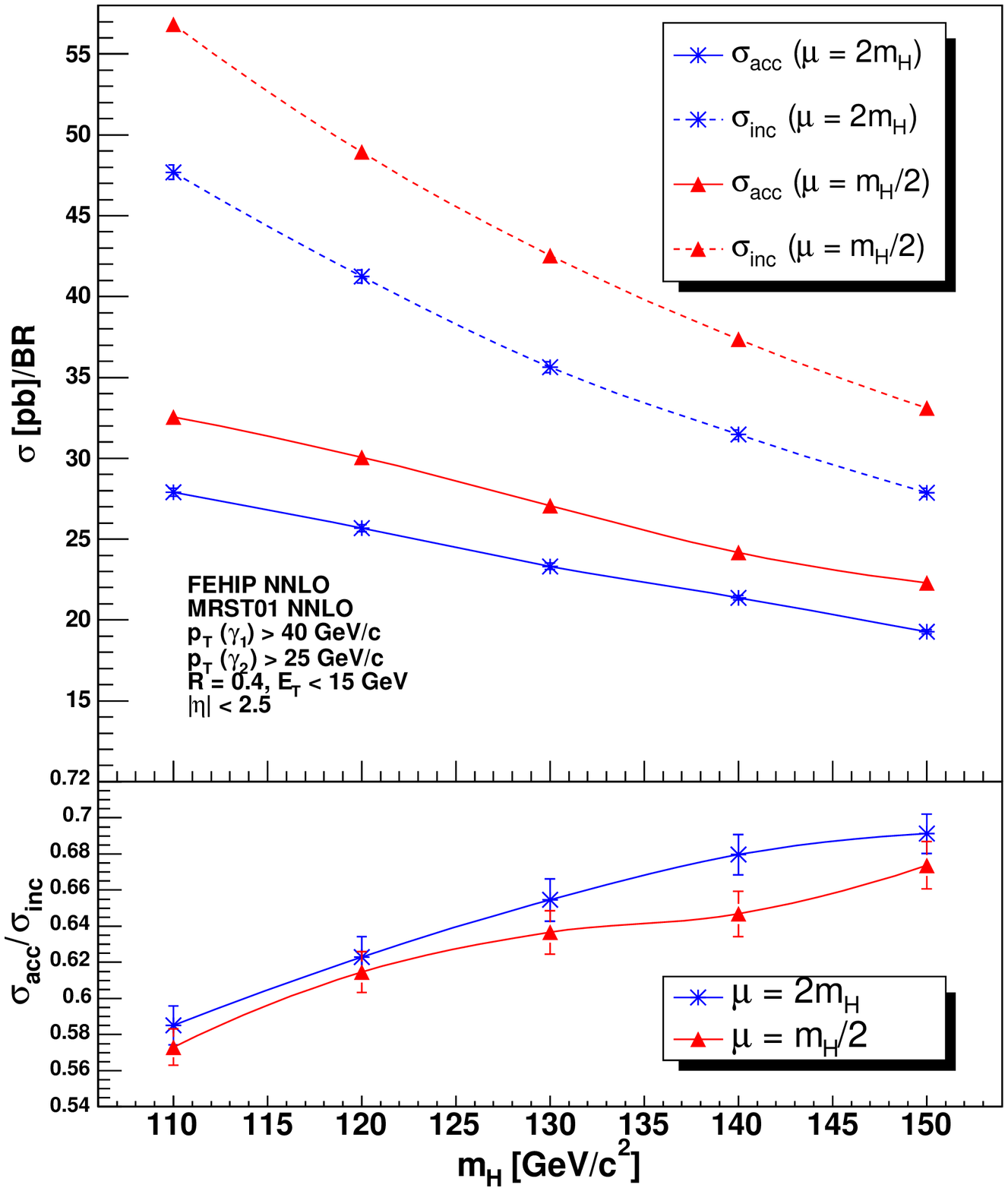}
  \caption{Inclusive and accepted cross sections at NNLO as a function of the
    Higgs mass, \mh, for two different scale choices.
    \label{fig:acc_cross_mu}
  }
\end{figure}

Both the renormalization and factorization scales are un-physical variables
which are introduced in the regularization step of the renormalization and/or
factorization procedure,  which is needed
 to deal with the divergent integrals encountered
in the perturbative calculations. An approximation for a perturbatively calculated
observable, obtained to some order in the coupling constant, does not depend
on these scales up to the calculated order, but a dependence appears at the
next-higher order. It is thus interesting to investigate the remaining dependence
of a given higher-order calculation. The range over which these scales are varied,
and the usually quoted systematic uncertainties corresponding to the spread in the
results when varying the scales, are somehow arbitrary and subject to controversy. 
For this analysis we have studied scale variations between $\mh/4$ and $2\,\mh$,
in both \MCaNLO\ and \FEHiP. It is worth noting that other settings of the scales
are possible in \MCaNLO, for example, including some \pt-dependence. Our approach
is convenient for a direct comparison of \MCaNLO\ and \FEHiP. 

\begin{figure}[htb]
  \centering
  \includegraphics[scale=0.7]{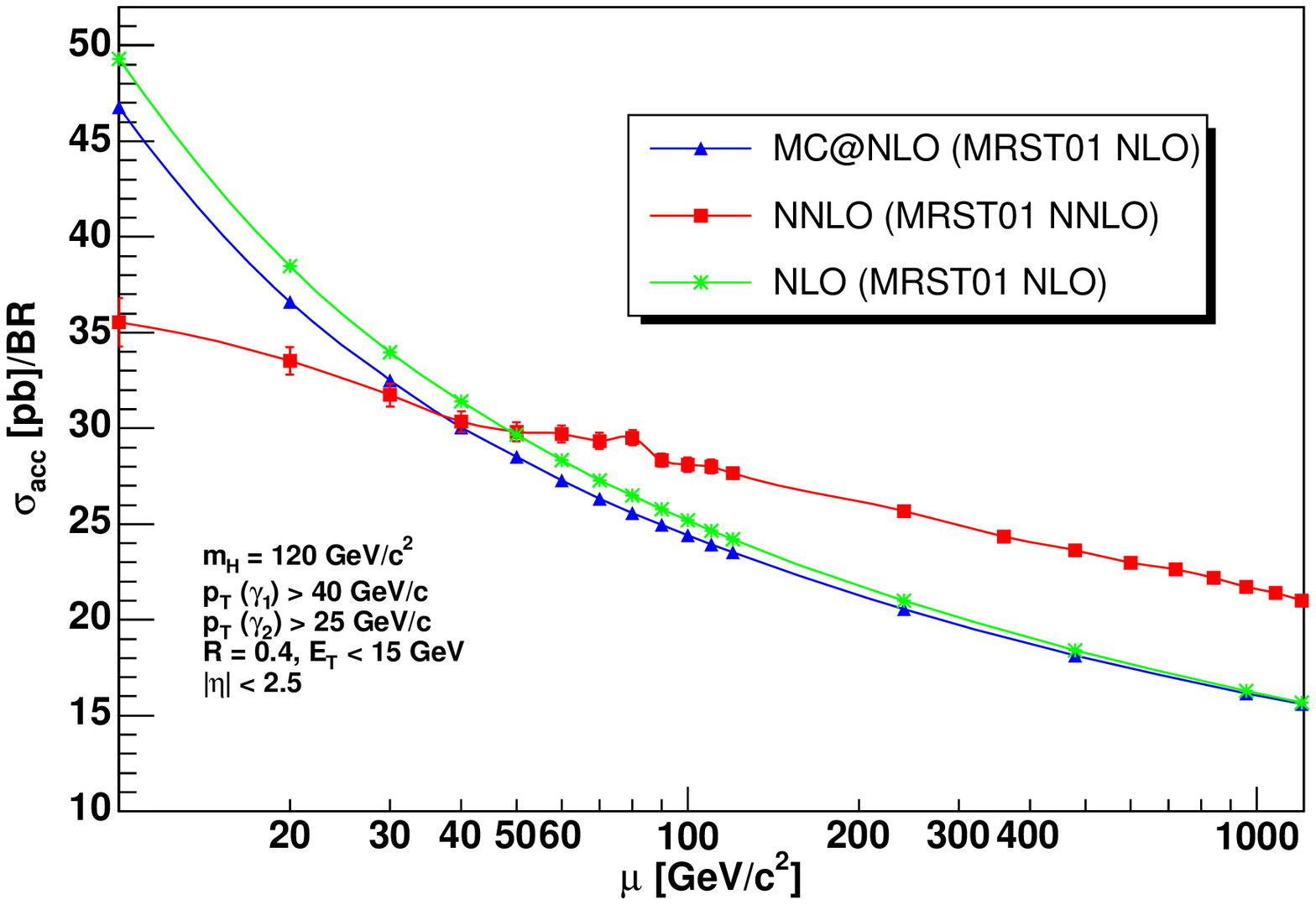}
  \caption{The accepted cross section \sacc\ predicted by \MCaNLO\ and \FEHiP\ (NLO and NNLO)
    as a function of the renormalization and factorization scale, $\mu$, for a Higgs mass of
     $\mh=120 \gevc2$. 
    \label{fig:scale-dep}
} 
\end{figure}

\begin{table}[htbp]
  \centering
  \begin{tabular}{|l|c|c|c|c|c|} \hline
    scale $\mu=\mu_{\mathrm{F}}=\mu_{\mathrm{R}}$ & $\mu=m_{\mathrm{h}}/4$ &  $\mu=m_{\mathrm{h}}/3$ & $\mu=m_{\mathrm{h}}/2$ & $\mu=m_{\mathrm{h}}$ & $\mu=2 m_{\mathrm{h}}$ \\ \hline
    $\sinc$ [pb]                         &  51.73     &  47.85     &  43.30     &  37.16     &  32.42              \\ \hline      
    \multicolumn{6}{c}{}                                                                                           \\ \hline
    \multicolumn{6}{|c|}{Acceptances ($\sigma_{\mathrm{acc}}/\sigma_{\mathrm{inc}}$) of single groups of cuts}     \\ \hline
     $p_{\mathrm{T}}$-cuts               &  80.2 \%   &  80.2 \%   &  80.2 \%   &  80.2 \%   &  80.1 \%            \\     
    $\eta$-cuts                          &  83.2 \%   &  83.2 \%   &  83.2 \%   &  83.4 \%   &  83.5 \%            \\ 
    isolation                            &  99.7 \%   &  99.7 \%   &  99.7 \%   &  99.8 \%   &  99.8 \%            \\ \hline
    \multicolumn{6}{c}{}                                                                                           \\ \hline
    \multicolumn{6}{|c|}{Acceptances ($\sigma_{\mathrm{acc}}/\sigma_{\mathrm{inc}}$) of two groups of cuts}        \\ \hline
    $p_{\mathrm{T}}$- and $\eta$-cuts    &  63.5 \%   &  63.4 \%   &  63.6 \%   &  63.8 \%   &  63.8 \%            \\ 
    $p_{\mathrm{T}}$-cuts and isolation  &  79.6 \%   &  79.6 \%   &  79.6 \%   &  79.7 \%   &  79.6 \%            \\ 
    $\eta$-cuts and  isolation           &  81.3 \%   &  81.3 \%   &  81.5 \%   &  81.7 \%   &  81.9 \%            \\ \hline 
    \multicolumn{6}{c}{}                                                                                           \\ \hline
    \multicolumn{6}{|c|}{All three groups of cuts}                                                                 \\ \hline
    Acceptance                           &  62.9 \%   &  62.8 \%   &  63.0 \%   &  63.3 \%   &  63.4 \%            \\ \hline
    $\sigma_{\mathrm{acc}}$ [pb]         &  32.52     &  30.05     &  27.30     &  23.53     &  20.54              \\ \hline
  \end{tabular}
  \caption{Acceptance corrections for different cuts and scale choices, as obtained with \MCaNLO. The Higgs mass is set to $\mh=120 \gevc2$. 
    \label{tab:higgs_scale_cuts}}
  
\end{table}

The NNLO results when choosing either $\mu = \mh/2$ or $\mu = 2\,\mh$ are shown in 
Fig.~\ref{fig:acc_cross_mu}.
As can be observed in Fig.~\ref{fig:scale-dep}, the NLO cross sections start
to increase very rapidly when going to very small scales, whereas the scale dependence
at NNLO is considerably flatter.
The difference between the NLO predictions from \FEHiP\ and \MCaNLO\ 
can be explained by the different treatment of the top mass dependence.

For a Higgs mass of $\mh=120 \gevc2$, as used in
Fig.~\ref{fig:scale-dep}, it appears that a range of scales between $\mh/2$ and $2\,\mh$
might be reasonable for evaluating a possible systematic uncertainty because of this effect.
Table~\ref{tab:higgs_scale_cuts} summarizes the \MCaNLO\ results  for the different scale
choices. Despite the rather large variation of the
cross sections when varying $\mu$, the acceptance correction is
remarkably stable to within $0.5\%$ (absolute), even when going to scales as low as $\mh/4$.
A similar stability of the acceptance correction 
has been found for single W production at LHC \cite{Frixione:2004us}.


%% file: diff_cross.tex
After having calculated the acceptance for a rather standard
set of basic selection cuts, it is interesting to investigate further
kinematic observables which might help to discriminate the signal
from backgrounds. In order to be able to optimize such an additional
selection, it is necessary to have a good understanding of the relevant 
kinematic distributions. 
Here we concentrate on two observables, which have also been discussed
in Refs.~\cite{Bern:2002jx,Anastasiou:2005qj}, namely the
distributions of the mean of the transverse momenta of the two 
final state photons $p_{\mathrm{T}}^{\mathrm{m}} =
\left(p_{\mathrm{T}}^{\gamma_1}+p_{\mathrm{T}}^{\gamma_2}\right)/2$
and the difference of their pseudo-rapidities
$\ys=\left|\eta^{\gamma_1}-\eta^{\gamma_2}\right|/2$.
As shown in \cite{Bern:2002jx}, the latter is interesting since a similar distribution
for the prompt photon  background is flatter. In the following
we focus on a Higgs mass of $\mh=120 \gevc2$ and set the renormalization
and factorization scale to $\mu=\mh/2$. All the distributions are
obtained
after applying the selection cuts as described in Section~\ref{sec:selection}.
 
\begin{figure}[htb]
  \centering
  \includegraphics[scale=0.8]{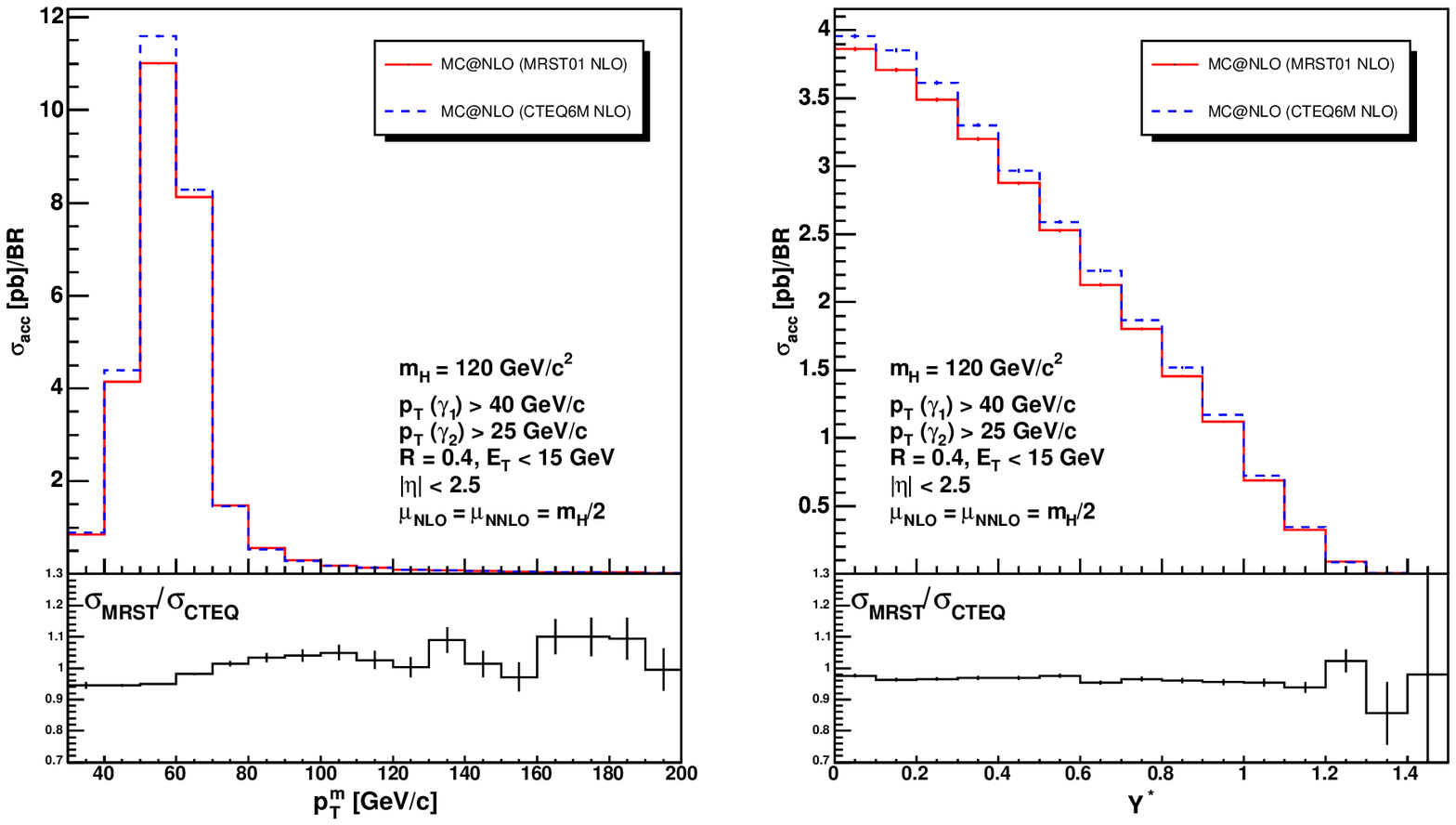}
  \caption{The distributions of the kinematic variables \ptm{} (left) and \ys{} (right), obtained with \MCaNLO\ for the
    two pdf sets \MRST\ NLO and \CTEQ\ NLO. 
    The standard cuts as indicated in the plots have been applied.
    \label{fig:pt_Y_cteq}
  }
\end{figure}

Figure~\ref{fig:pt_Y_cteq}  shows the \ptm{} (left) and \ys{} (right) distributions 
obtained with \MCaNLO\ when using two different pdf sets, \MRST\ NLO and 
\CTEQ\ NLO. We observe that in the interesting region $30\gevoc \le
\ptm \le 100\gevoc$ the differential cross section found with  \CTEQ\  exceeds that
obtained with the \MRST\ NLO set by up to 5\%, in accordance with
the differences seen for the inclusive and accepted cross sections listed in 
Table~\ref{tab:higgs-efficiencies-cuts}.  Obviously,
the same feature can be found in the \ys{} distribution. However, in this case it is
remarkable that the ratio of the two predictions is flat over the whole \ys\ range,
which indicates that the shape of this
distribution is rather insensitive to the choice of the pdf set. 

\begin{figure}[htb]
  \centering
  \includegraphics[scale=0.8]{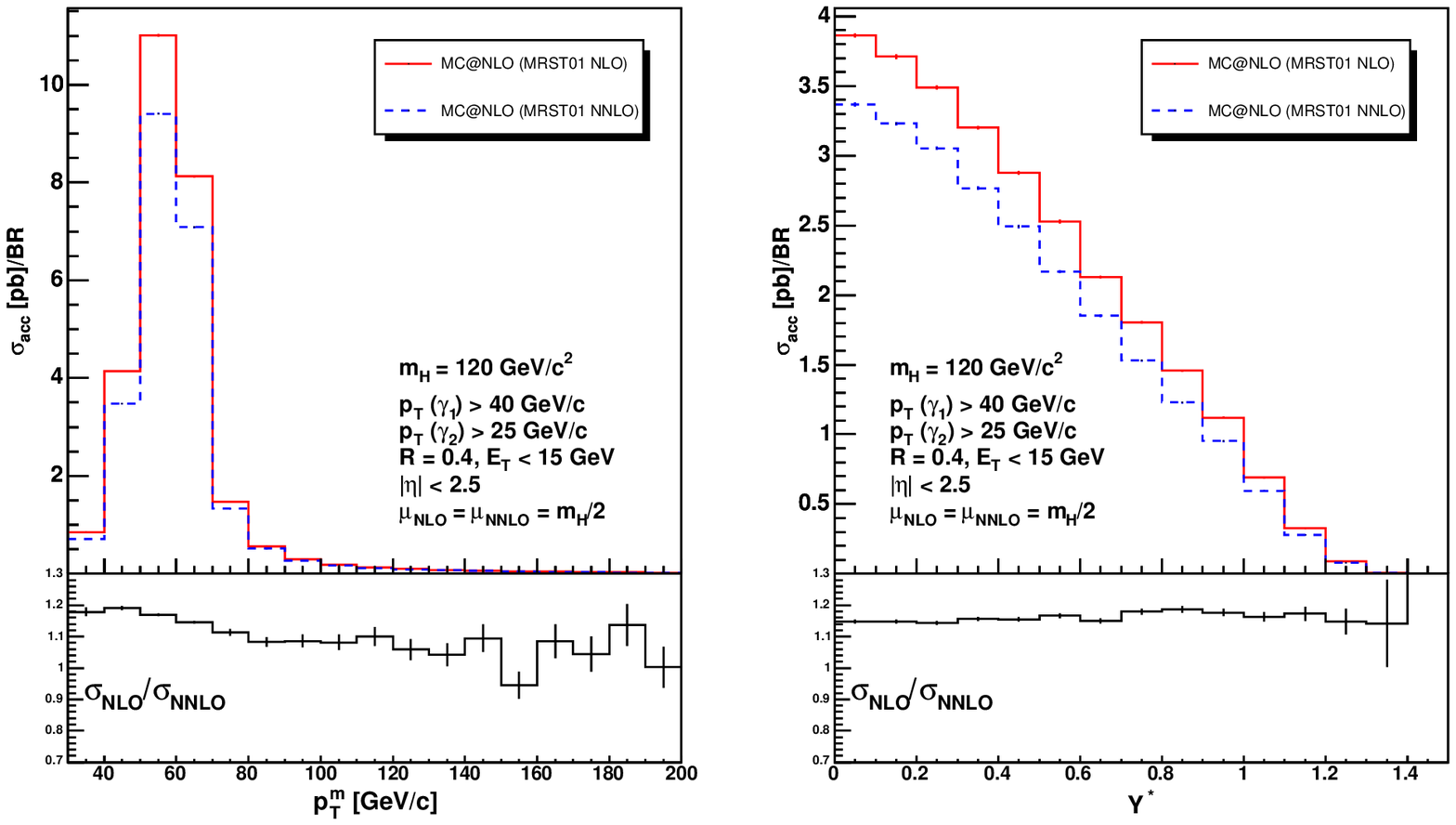}
  \caption{The distributions of the kinematic variables \ptm{} (left) and \ys{} (right),
     obtained with \MCaNLO\ for the
    two pdf sets \MRST\ NLO and \MRST\ NNLO.
    The standard cuts as indicated in the plots have been applied.
    \label{fig:pt_Y_mrstnnlo}
  }
\end{figure}

The same distributions are reproduced in Fig.~\ref{fig:pt_Y_mrstnnlo}, 
where now the \CTEQ\ pdfs have been replaced by those from the \MRST\ NNLO set. 
Similarly to the observations above, the shape of the two 
histograms differ only slightly, in particular in the case of \ys, 
with the results based on the NNLO pdfs being lower than those from
the NLO set. This has been discussed previously, cf.\ Section \ref{pdf-inc} and 
Table~\ref{tab:higgs-efficiencies-cuts}. 
Thus the assumption is confirmed that the choice of the pdf set does not strongly  
affect the shape of these distributions.

\begin{figure}[htb]
  \centering
  \includegraphics[scale=0.8]{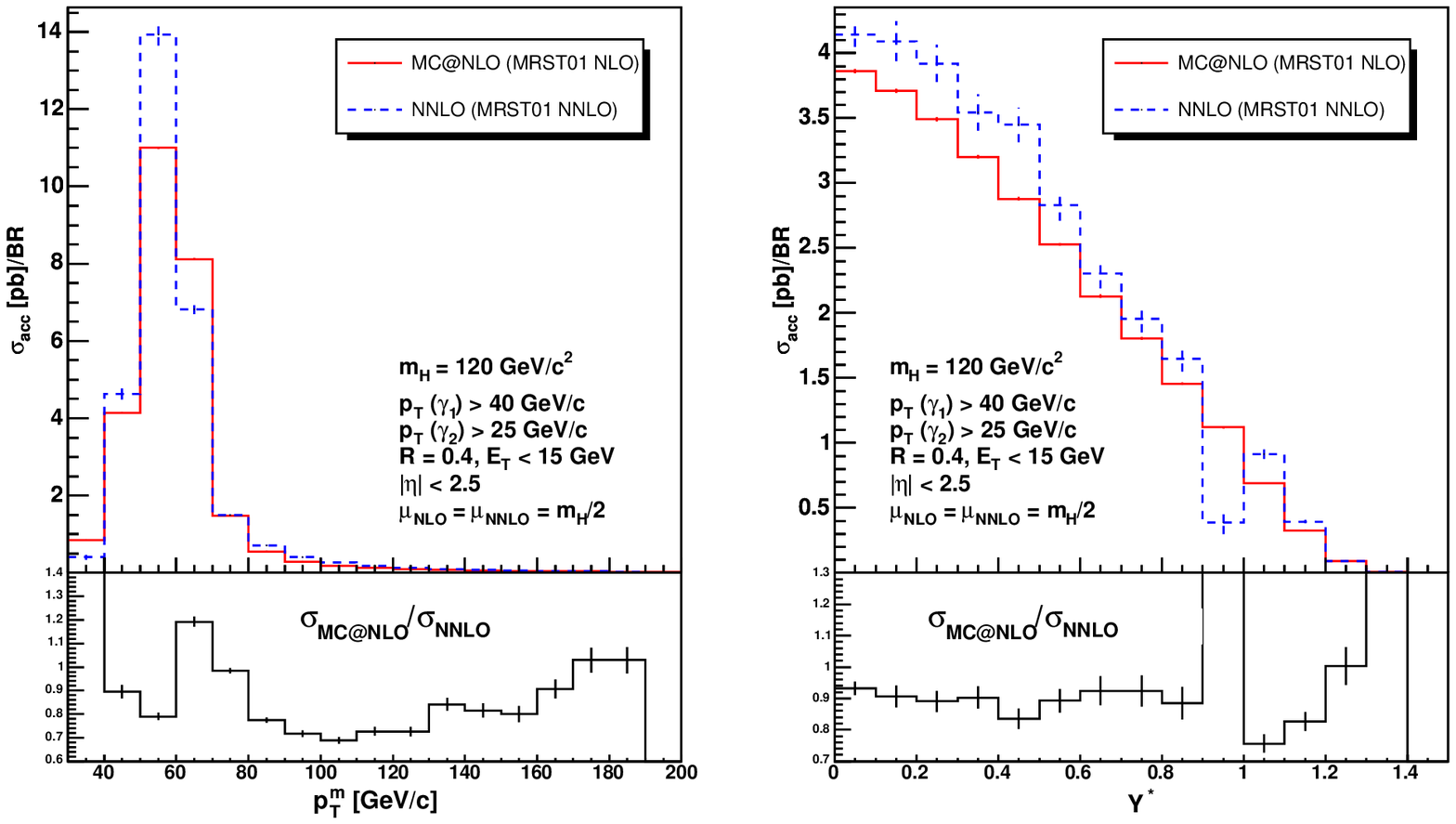}
  \caption{The distributions of the kinematic variables \ptm{} (left) and \ys{} (right),
    obtained with \MCaNLO\ (pdf set \MRST\ NLO) and \FEHiP\ (NNLO).
    The standard cuts as indicated in the plots have been applied.
    \label{fig:pt_Y_nnlo}
 }
\end{figure}

Finally, in Fig.~\ref{fig:pt_Y_nnlo} we compare the distributions obtained with \MCaNLO\
and with \FEHiP\ at NNLO. Concerning the variable
\ptm{},   we observe that in the peak region the shapes of the distributions 
 differ rather strongly, with the maximum
shifted to lower values when going from \MCaNLO\ to NNLO. The
appearance of these large perturbative corrections close to the kinematic
boundary has also been discussed in \cite{Anastasiou:2005qj}. 
From the upper tail, thus far beyond this boundary, it is evident that
at larger photon momenta the cross section is higher at NNLO with respect to \MCaNLO.
Regarding the \ys{} distribution it is particularly interesting to observe that
the step in the NNLO prediction around $\ys=0.9-1.0$ does not appear
for \MCaNLO.  Such a step is typical for fixed order calculations when
a phase-space boundary varies
at different orders in the perturbation series, however, it can be removed by the inclusion
of soft-gluon resummation to all orders~\cite{Catani:1997xc}. In the case of \ys{}, the leading order
cross section is zero above $\ys=0.96$.

%% file: background.tex
After having analyzed the signal cross sections as functions of the 
measurable final state photon momenta,  it is natural to compare these
differential results to those for the irreducible background, namely
prompt photon production, \pp \ra  \ggpX. The latter have been computed
with \DiPHOX \cite{Binoth:1999qq} at NLO (cf.\ Section \ref{sec:programs}),
with \MRST\ NLO as pdf set. The renormalization and factorization
scales are fixed to $\mu=\mh/2$ for the signal, computed with \MCaNLO, and 
$\mu=m_{\gg}/2$ for the background, where $m_{\gg}$ is the invariant 
mass of the two final-state photons. For the photon fragmentation we 
use the NLO  fragmentation functions (set I) from \cite{Bourhis:1999}.

\begin{figure}[htb]
  \centering
  \includegraphics[scale=0.8]{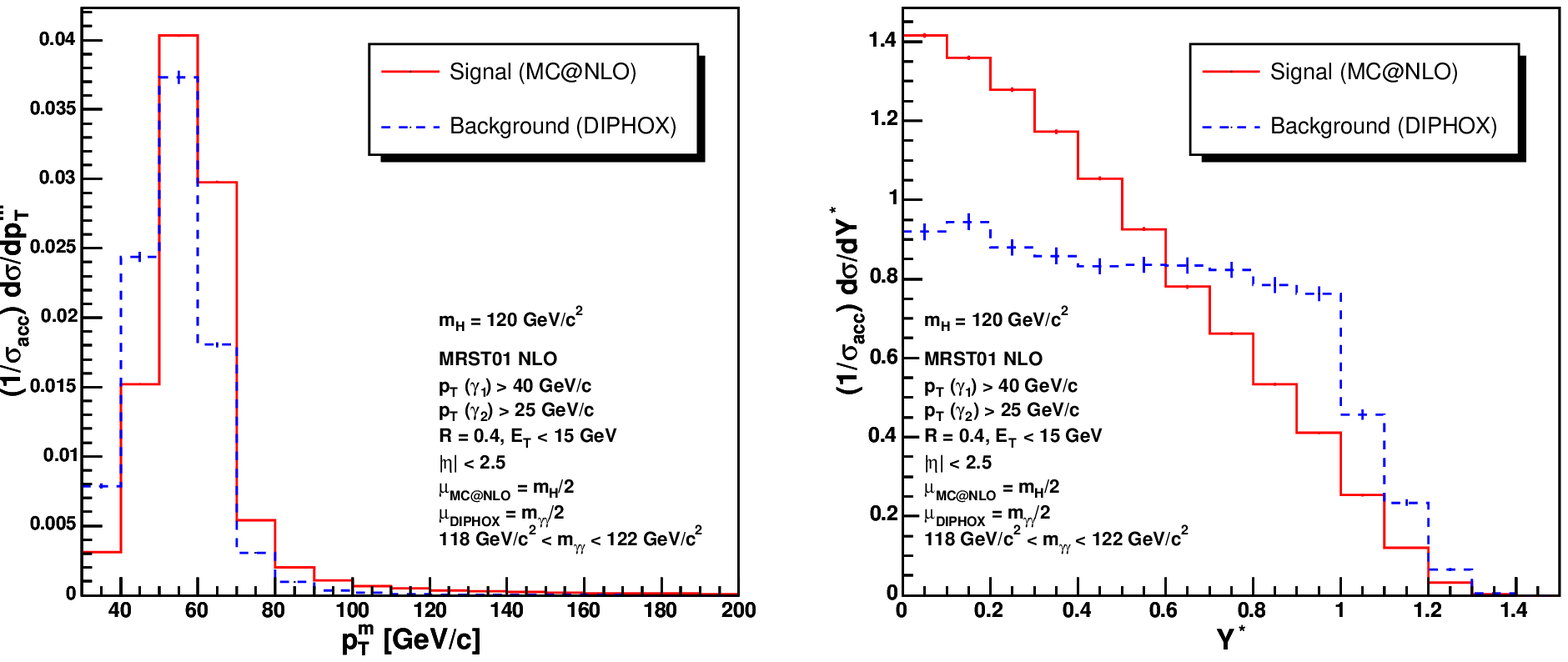}
  \caption{Normalized distributions of \ys{} and \ptm{} for signal and background.
                 The standard cuts as indicated in the plots have been applied.
\label{fig:sig_back}}

\end{figure}

Figure~\ref{fig:sig_back} compares the normalized 
distributions of \ptm{} (left) and \ys\ (right) for signal and background.
 For the background, only events 
with $118 \gevc2 \le m_{\gg} \le 122 \gevc2$ are included, 
corresponding to a mass-window of $\pm 2 \gevc2$ around the Higgs mass $\mh=120 \gevc2$.
As expected, the shapes differ considerably. In the case of \ptm, the mean
is clearly shifted towards lower  values for the background. The \ys\ distribution
offers an even better handle for the signal-background discrimination, since the background
is rather flat, whereas the signal is peaked for small values of \ys.

\subsection{Statistical significance}

In Table~\ref{tab:significance} we compare the signal 
cross section obtained with \MCaNLO\ to  the background cross section computed  
with \DiPHOX. Here, \sinc{} refers to the inclusive 
\mbox{\pp\ra\H\ra\ggpX} cross section before any cut. In addition to the 
acceptance correction found with \MCaNLO\ (cf.\ Section \ref{sec:incxsect}), we also
account for a photon reconstruction efficiency of $57~\%$ 
as was done in Ref.~\cite{Bern:2002jx}, 
corresponding to a combination of $81\%$ for $\gamma$/jet identification 
per photon and $87\%$ for fiducial cuts. 
The branching ratio for the Higgs decay channel into two photons is 
calculated with \HDECAY. Altogether this gives an effective signal cross 
section of $\sigma_{\mathrm{eff}}=0.034\,\mathrm{pb}$. If we assume an 
integrated luminosity of $30\,\mathrm{fb}^{-1}$, corresponding to about
three years of \LHC\ running at low luminosity, we find a total number of $S=1027$ signal 
events.

In order to calculate the number of background events, we only consider events with a total
invariant mass of the two final-state photons between $118$ and $122\gevc2$. In this case, the 
cross section after cuts amounts to $\sacc=0.944$ pb, as shown in the right column of
Table~\ref{tab:significance}. In addition  to the efficiency factor, we 
include a $20~\%$ reducible background \cite{Bern:2002jx}.
This leads to an effective background cross section of 
$\sigma_{\mathrm{eff}}=0.631\;\mathrm{pb}$ and to a total number of 
$B=18928$ background events. Thus, we end up with a significance of $S/\sqrt{B} = 7.5$.

A few remarks are in place here. First, this is the significance obtained only after
applying the basic selection cuts given in Section \ref{sec:selection}, without any further
signal-background discrimination as suggested, eg.\ by the observations 
from Fig.~\ref{fig:sig_back}. Second, \DiPHOX\ does not contain the two-loop correction to the gluon fusion
subprocess of \pp \ra \ggpX, which has been calculated in  \cite{Bern:2002jx}.
However, its inclusion would only lead to a 5\%  reduction of the statistical
significance given in Table~\ref{tab:significance}.
Furthermore, it is worth noting
that the background has been calculated at NLO, while
the signal obtained with \MCaNLO\ includes corrections beyond that order. We think that a
calculation of the prompt photon background to such an approximation as can be
obtained with the \MCaNLO\ approach would be highly desirable, both from the
phenomenological and experimental point of view.

\begin{table}[htbp]
  \centering
  \begin{tabular}{|l|c|l|c|} \hline
    \multicolumn{4}{|c|}{\pp \ra \ggpX}                                                                      \\ \hline
    \multicolumn{2}{|c|}{Signal}                        & \multicolumn{2}{|c|}{Background}                   \\ \hline
    $\sigma_{\mathrm{inc}}$ [pb]           & 43.30      &                                      &             \\ 
    acceptance                             & 63.0 \%    &                                      &             \\ 
    $\sigma_{\mathrm{acc}}$ [pb]           & 27.30      & $\sacc$ [pb]                         & 0.9244      \\ 
    efficiency factor                      & 57 \%      & efficiency factor                    & 57 \%       \\ 
    branching ratio                        & 0.22 \%    & other reducible background           & 20 \%       \\ 
    $\sigma_{\mathrm{eff}}$ [pb]           & 0.034      & $\sigma_{\mathrm{eff}}$ [pb]         & 0.631       \\ 
    int.{} luminosity $[\mathrm{fb}^{-1}]$ & 30         & int. luminosity $[\mathrm{fb}^{-1}]$ & 30          \\ 
    number of events $S$                   & 1027       & number of events $B$                 & 18928       \\ \hline
    \multicolumn{4}{|c|}{}                                                                                   \\
    \multicolumn{4}{|c|}{significance $S/\sqrt{B} = 7.5$ }                                                   \\\hline 
  \end{tabular}
  \caption{\label{tab:significance} Accepted number of signal and
    background events after applying standard selection cuts and corresponding 
    significance $S/\sqrt{B}$ for a Higgs mass of $\mh=120\gevc2$.}
\end{table}


%% file: summary.tex
We have analyzed perturbative QCD predictions at and beyond 
next-to-leading order accuracy for the Higgs production and
decay in the channel  \pp \ra \H \ra \ggpX\ at \LHC\ energies.
In particular, for the first time it has become possible to compare 
the results from the \MCaNLO\ Monte Carlo generator to a full
NNLO prediction, not only for inclusive cross sections, but also
for the accepted cross sections 
obtained after applying standard selection cuts.

Whereas the inclusive and accepted cross sections at NLO and NNLO differ by about 10\%,
it turns out that the acceptance corrections agree to within 2\%. The latter
are also very stable under variations of the renormalization and factorization scale and rather
insensitive to the choice of the parton distribution functions. This is 
not the case when looking at the cross section themselves, in particular at NLO approximation.

We have also analyzed distributions of kinematic observables constructed
from the final-state photon momenta, namely the average of the two
photon transverse momenta and their rapidity difference.
Some differences in the shapes of the distributions have been
found between \MCaNLO\ and the exact NNLO calculation. More importantly,
it has been shown that these distributions should allow for a 
good discrimination between the Higgs signal and the irreducible
prompt photon background.
